\documentclass[12pt]{article}
\usepackage{ada-grover}
\begin{document}

\title{An Adaptive, Fixed-Point\\
Version of Grover's Algorithm}

\author{Robert R. Tucci\\
        P.O. Box 226\\
        Bedford,  MA   01730\\
        tucci@ar-tiste.com}

\date{ \today}

\maketitle

\vskip2cm
\section*{Abstract}
We give an adaptive,
fixed-point version of
Grover's algorithm.
By this we mean
that our algorithm
performs an infinite sequence of
gradually diminishing steps (so
we say it's adaptive) that
drives
the starting
state to the target
state
with absolute certainty
(so we say it's a fixed-point algorithm).
Our algorithm
is motivated by Bloch sphere
geometry. We include with
the ArXiv distribution of
this paper some
simple software (Octave/Matlab
m-files) that
implements,
tests and illustrates some of the
results of this paper.

\newpage
\section{Introduction}

Grover proposed his
original algorithm in Ref.\cite{Grover-model-O}.
His algorithm takes a starting
state towards
a target state by
performing a sequence of equal steps.
By this we mean that
each step is
a rotation about the same fixed axis
and
by the same  small angle.
Because each step is
by the same angle,
the algorithm overshoots
past the target state
once it reaches it.
Grover later proposed
in Ref.\cite{Grover-model-F}
a ``$\pi/3$ fixed-point"
algorithm which uses
a recursion relation
to define an infinite sequence of
gradually diminishing steps that
drives
the starting
state to the target
state
with absolute certainty.

Other workers  have
pursued what they
refer to as a phase matching approach
to Grover's algorithm.
Ref.\cite{Toy} by Toyama et al.
is a recent
contribution to that approach,
and has a very complete review of previous
related contributions.

In this paper,
we describe what we call
an Adaptive, Fixed-point,
Grover's Algorithm (AFGA, like
Afgha-nistan, but without the h).
Our AFGA is motivated by Bloch sphere
geometry.
Our AFGA
resembles the original
Grover's algorithm
in that it applies
a sequence of rotations
about the same fixed axis, but
differs from it in that
the angle of successive rotations is different.
Thus, unlike the original Grover's algorithm,
our AFGA performs a sequence of
unequal steps. Our AFGA resembles
Grover's $\pi/3$ algorithm in
that it is a fixed-point
algorithm that converges to the
target, but it differs
from the $\pi/3$ algorithm
in its choice
of sequence of unequal steps.
Our AFGA resembles the
phase-matching approach
of Toyama et al., but their
algorithm uses only
a finite number of distinct
``phases",
whereas our AFGA uses an
infinite number. The
Toyama et al. algorithm
is not guaranteed to
converge to the target
(in the single-target case,
which is what concerns us in this paper),
so, it is not a true fixed-point
algorithm.

This paper was born
as an attempt to fill
a gap in a previous paper, Ref.\cite{TucGibbs},
which proposes a quantum Gibbs sampling
algorithm. The algorithm of
Ref.\cite{TucGibbs} requires
a version of Grover's
algorithm that works even if
there is a large
overlap between the starting
state and the target state. The original
Grover's algorithm only works
properly if that overlap
is very small. This
paper gives a version of Grover's
algorithm without the small overlap limitation.

We include with
the ArXiv distribution of
this paper some
simple software (Octave/Matlab
m-files) that
implements,
tests and illustrates some of the
results of this paper.
(Octave is a freeware partial
clone of Matlab. Octave m-files
should run in the Matlab
environment with zero or few modifications.)

Some vocabulary and notational conventions
that will be used in this paper:
Let $\ns = 2^\nb$ be the number of states
for $\nb$ bits.
Unit $\RR^3$ vectors will be denoted by letters with
a caret over them.
For instance, $\hatr$.
We will often abbreviate $\cos(\theta)$
and $\sin(\theta)$, for some $\theta\in \RR$,
by $C_\theta$ and $S_\theta$.

We will say a problem
can be solved {\bf with
polynomial efficiency}, or {\bf
p-efficiently} for short, if
its solution can be
obtained in a
time polynomial in $\nb$. Here $\nb$ is
the number of bits
required to encode the
input for the algorithm
that solves the problem.

By {\bf compiling a unitary matrix},
we mean decomposing it into a {\bf SEO}
(Sequence of
Elementary Operations).
Elementary operations
are one or two qubit operations such as
single-qubit rotations and CNOTs.
Compilations can be either
exact, or approximate (within
a certain precision).

We will say a unitary operator $U$
acting on $\CC^\ns$ can be
{\bf compiled with polynomial efficiently},
or {\bf p-compiled} for short,
if $U$
can be expressed,
either approximately
or exactly,
as a SEO
whose number of elementary operations (the SEOs length)
is polynomial in $\nb$.

\section{Review of Pauli Matrix Algebra}

The Pauli matrices are

\beq
\sigx =
\left[
\begin{array}{cc}
0&1\\1&0
\end{array}
\right],\;\;
\sigy =
\left[
\begin{array}{cc}
0&-i\\i&0
\end{array}
\right],\;\;
\sigz =
\left[
\begin{array}{cc}
1&0\\0&-1
\end{array}
\right]
\;.
\eeq

Define
\beq
\vec{\sigma} = (\sigx,\sigy,\sigz)
\;.
\eeq
For any $\vec{a}\in \RR^3$, let
\beq
\sigma_{\vec{a}} = \vec{a}\cdot\vec{\sigma}
\;.
\eeq
(I often refer to $\sigma_{\vec{a}}$ as
a ``Paulion", related to a quaternion).

Suppose $\vec{a}, \vec{b}\in \RR^3$.
The well know property of Pauli Matrices
\beq
\sigma_p\sigma_q = \delta_{p,q} + i \epsilon_{pqr}\sigma_r
\;,
\eeq
where indices $p,q,r$ range over $\{1,2,3\}$,
immediately implies that
\beq
\sigma_{\vec{a}} \sigma_{\vec{b}} =
\vec{a}\cdot\vec{b}
+ i\sigma_{\vec{a}\times{\vec{b}}}
\;.
\eeq

\begin{figure}[h]
    \begin{center}
    \epsfig{file=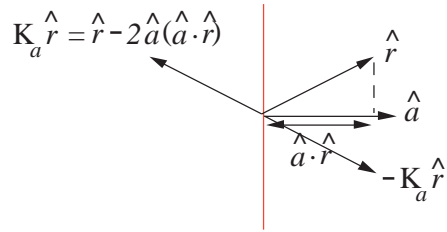, height=1.2in}
    \caption{$K_\hata\hatr$ is the reflection
    of $\hatr$ on the plane perpendicular
    to $\hata$.}
    \label{fig-reflection}
    \end{center}
\end{figure}

For any unit $\RR^3$ vectors $\hatr,\hata$, define
\beq
K_{\hata}\hatr = \hatr - 2 \hata(\hata\cdot\hatr)
\;.
\eeq
From Fig.\ref{fig-reflection},
we see that $K_{\hata}$
is a reflection: it
reflects the vector $\hatr$ on
the plane perpendicular to $\hata$.
Furthermore,
$-K_{\hata}$
is a pi rotation: it
rotates the vector $\hatr$
by an angle of pi, about the
axis $\hata$. Note that

\beqa
\sigma_\hata \sigma_\hatr \sigma_\hata
&=&
(\hata\cdot\hatr + i \sigma_{\hata\times\hatr})\sigma_\hata
\\ &=&
(\hata\cdot\hatr)\sigma_\hata -
\sigma_{(\hata\times\hatr)\times\hata}
\\ &=&
(\hata\cdot\hatr)\sigma_\hata -
\sigma_{\hatr - \hata(\hata\cdot\hatr)}
\\&=&
\sigma_{-K_{\hata}\hatr}
\;.
\eeqa

\begin{figure}[h]
    \begin{center}
    \epsfig{file=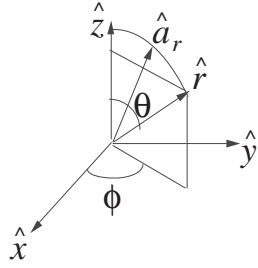, height=1.6in}
    \caption{A pi rotation of $\hatz$
   about $\hata_r$ yields $\hatr$.}
    \label{fig-a-r}
    \end{center}
\end{figure}

The eigenvectors of
$\sigz$ are given by

\beq
\sigz\ket{\pm \hatz} = (\pm 1) \ket{\pm \hatz}
\;,
\label{eq-evecs-sigz}
\eeq
where

\beq
\ket{+\hatz}=
\left[\begin{array}{c}
1\\0
\end{array}\right]=
\ket{0}
\;,\;\;
\ket{-\hatz}=
\left[\begin{array}{c}
0\\1
\end{array}\right]=
\ket{1}
\;.
\eeq
It is also convenient
to consider the eigenvectors
of $\sigma_\hatr$,
for any unit $\RR^3$ vector $\hatr$.
Given $\hatr$, we can
always find a unit $\RR^3$ vector $\hata_r$
such that
$\hatr$
is obtained by pi rotating
$\hatz$ about $\hata_r$. (See Fig.\ref{fig-a-r}).
Eq.(\ref{eq-evecs-sigz}) implies that
\beq
(\sigma_{\hata_r} \sigma_\hatz \sigma_{\hata_r})
\sigma_{\hata_r}\ket{0}=
\sigma_{\hata_r}\ket{0}
\;.
\label{eq-intermediate-evec-eq}
\eeq
Furthermore,

\beq
 \sigma_\hatr
 =
 \sigma_{\hata_r} \sigma_\hatz \sigma_{\hata_r}
\;.
\eeq
Hence, if we define $\ket{\hatr}$ by

\beq
\ket{\hatr} = \sigma_{\hata_r}\ket{0}
\;,
\label{eq-def-evec-sigr}
\eeq
the Eq.(\ref{eq-intermediate-evec-eq}) becomes

\beq
\sigma_{\hatr}\ket{\hatr} =  \ket{ \hatr}
\;.
\label{eq-plusr-evec-eqt}
\eeq
Eq.(\ref{eq-plusr-evec-eqt}) itself implies that

\beq
\sigma_{\hatr}\ket{\pm\hatr} =  \pm\ket{\pm \hatr}
\;.
\eeq
Thus, $\ket{\pm\hatr}$
defined by Eq.(\ref{eq-def-evec-sigr})
are the eigenvectors
of $\sigma_\hatr$ with
eigenvalues $\pm 1$.

Given $\hatr$ in polar coordinates,
we can express $\ket{\hatr}$
in terms of the polar coordinates of $\hatr$,
as follows. Suppose

\beq
\hatr =
\left[
\begin{array}{c}
\sin\theta\cos\phi\\
\sin\theta\sin\phi\\
\cos\theta
\end{array}
\right]
\;,
\eeq
where $\theta\in[0,\frac{\pi}{2}]$
and $\phi\in[0,2\pi)$.
Then, by definition, $\hata_r$
has the same $\phi$ but half the $\theta$
coordinate:

\beq
\hata_r =
\left[
\begin{array}{c}
\sin(\frac{\theta}{2})\cos\phi\\
\sin(\frac{\theta}{2})\sin\phi\\
\cos(\frac{\theta}{2})
\end{array}
\right]
\;.
\eeq
Thus,

\beq
\ket{\hatr}=
\sigma_{\hata_r}\ket{0}=
\left[\begin{array}{c}
\hata_z\\ \hata_x + i \hata_y
\end{array}\right]=
\left[\begin{array}{c}
\cos(\frac{\theta}{2})\\
e^{i\phi}\sin(\frac{\theta}{2})
\end{array}\right]
\;.
\label{eq-ketr-polar}
\eeq
From Eq.(\ref{eq-ketr-polar}),
we get the following table.

\beq
\begin{array}{c|c}
\RR^3 & \CC^2
\\ \hline
\pm \hatx & \ket{\pm \hatx}=
\frac{1}{\sqrt{2}}\left[
\begin{array}{c}
1\\ \pm 1
\end{array}
\right]
\\\hline
\pm \haty & \ket{\pm \haty}=
\frac{1}{\sqrt{2}}\left[
\begin{array}{c}
1\\ \pm i
\end{array}
\right]
\\\hline
\hatz,-\hatz &
\ket{\hatz}=
\left[
\begin{array}{c}
1\\ 0
\end{array}
\right],
\ket{-\hatz}=
\left[
\begin{array}{c}
0\\ 1
\end{array}
\right]
\end{array}
\;
\eeq
In general, the
assignment $\hatr\mapsto\ket{\hatr}$
yields a map of $\RR^3$ into
$\CC^2$, the 2-dimensional
complex vector space
spanned by complex
linear combinations of $\ket{0}$
and $\ket{1}$.
If we confine ourselves to the
$\hatx-\hatz$ plane of $\RR^3$,
then that plane is mapped
into the {\it half}-plane of
all {\it real} linear combinations
of $\ket{0}$ and $\ket{1}$
with non-negative $\ket{0}$ component.
(Kets that differ
by a phase factor or
a normalization constant
are equivalent).

Just like it
is useful to consider the projection
operators $\ket{0}\bra{0}$
and $\ket{1}\bra{1}$, it is
also useful to consider the projection
operator $\ket{\hatr}\bra{\hatr}$.
Recall the usual definitions
of the number operator $n$ and
its complement $\nbar$:
\begin{subequations}\label{eq-p0-p1}
\beq
n = P_1=\ket{1}\bra{1}= \frac{1-\sigz}{2}
\;,
\eeq

\beq
\nbar = 1-n = P_0=
\ket{0}\bra{0}=
\frac{1 + \sigz}{2}
\;.
\eeq
\end{subequations}
Rotating the coordinate system
so that $\hatz$ goes to $\hatr$, we get
\begin{subequations}
\beq
n_\hatr = P_{-\hatr}=
\ket{-\hatr}\bra{-\hatr} =
\frac{1-\sigma_\hatr}{2}
\;,
\eeq

\beq
\nbar_\hatr = 1 - n_\hatr=
P_{\hatr}=
\ket{\hatr}\bra{\hatr} =
\frac{1+\sigma_\hatr}{2}
\;.
\eeq
\end{subequations}
Note that if we define
the reflection operator
$K_{\ket{\hatr}}$ by

\beq
K_{\ket{\hatr}}= 1 - 2\ket{\hatr}\bra{\hatr}
\;,
\eeq
then

\beq
K_{\ket{\hatr}}= 1 -2 \nbar_\hatr=(-1)^{\nbar_\hatr}
=-\sigma_\hatr
\;.
\eeq

In general,
the vectors in
$\CC^2$ are
``packed twice as densely"
as the corresponding
vectors in $\RR^3$.
By this we mean that the angle between
two vectors in $\CC^2$
is always half the angle
between
the corresponding vectors in
$\RR^3$. To prove this, note that

\beqa
|\av{\hatr_1|\hatr_2}|^2 &=&
\tr(\ket{\hatr_1}\bra{\hatr_1}\ket{\hatr_2}\bra{\hatr_2})
\\&=&
\frac{1}{4}\tr[
(1+\sigma_{\hatr_1})
(1+\sigma_{\hatr_2})
]
\\&=&
\frac{1}{4}\tr(1 + \sigma_{\hatr_1}\sigma_{\hatr_2})
\\&=&
\frac{1}{4}
\tr(1 + \hatr_1\cdot\hatr_2)
\\&=&
\frac{1}{2}
(1 + \hatr_1\cdot\hatr_2)
\;.
\eeqa
Thus, if $\hatr_1\cdot\hatr_2=\cos\alpha$, then
$|\av{\hatr_1|\hatr_2}|
= \sqrt{\frac{1+\cos\alpha}{2}}
=|\cos(\frac{\alpha}{2})|$.

Suppose
$\theta\in\RR$.
A Taylor expansion easily establishes that
\beq
e^{i\theta\sigz}
= \cos\theta + i\sigz\sin\theta
\;.
\eeq
Rotating the coordinate system
so that $\hatz$ goes to $\hatr$, we get

\beq
e^{i\theta\sigma_\hatr}
= \cos\theta + i\sigma_\hatr\sin\theta
\;.
\eeq

Suppose that $\hata$ and
$\hatb$ are two unit $\RR^3$ vectors
which make an angle $\theta$
between them. Let
$\hatn = \frac{\hata\times\hatb}{|\hata\times\hatb|}$
be the unit $\RR^3$ vector normal to
the plane defined by $\hata$ and $\hatb$.
Then

\beq
\sigma_\hata \sigma_\hatb =
\hata\cdot\hatb
+ i\sigma_{\hata\times\hatb}
=\cos\theta +
i\sigma_\hatn
\sin\theta
= e^{i\theta\sigma_\hatn}
\;.
\eeq
Thus, any SU(2) element
$e^{i\theta\sigma_\hatn}$,
where $\theta\in\RR$ and
$\hatn$
is a unit $\RR^3$ vector, can
be expressed (non-uniquely)
as a product of
two Paulions $\sigma_\hata$
and $\sigma_\hatb$.

\begin{figure}[h]
    \begin{center}
    \epsfig{file=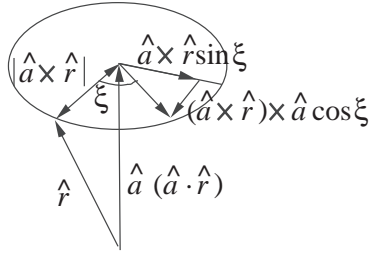, height=1.3in}
    \caption{A finite rotation of $\hatr$
    about axis $\hata$ by an angle of $\xi$.}
    \label{fig-full-rot}
    \end{center}
\end{figure}

From Fig.\ref{fig-full-rot},
it is clear that if
$R_{\hata}(\xi)$
is the rotation operator
that rotates any unit $\RR^3$ vector
$\hatr$, by an angle $\xi\in\RR$, about an axis
defined by the unit $\RR^3$ vector
$\hata$,
then

\beq
R_{\hata}(\xi)\hatr=
\hata(\hata\cdot\hatr)
+\sin(\xi)\hata\times\hatr
+\cos(\xi)[\hatr - \hata(\hata\cdot\hatr)]
\;.
\eeq

\begin{figure}[h]
    \begin{center}
    \epsfig{file=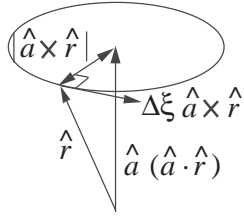, height=1.1in}
    \caption{An infinitesimal rotation of $\hatr$
    about axis $\hata$ by an angle of $\Delta\xi$.}
    \label{fig-small-rot}
    \end{center}
\end{figure}

If $\hatr,\hata$ are unit $\RR^3$ vectors
and $\Delta\xi$ is an infinitesimal
real number, then

\beq
e^{-i\frac{\Delta\xi}{2}\sigma_\hata}
\sigma_\hatr
e^{i\frac{\Delta\xi}{2}\sigma_\hata}
\approx
\sigma_\hatr -
i \frac{\Delta\xi}{2}(\sigma_\hata\sigma_\hatr
-\sigma_\hatr\sigma_\hata)
=
\sigma_{\hatr + \Delta\xi (\hata\times\hatr)}
\;.
\eeq
From Fig.\ref{fig-small-rot},
$\hatr + \Delta\xi (\hata\times\hatr)$
is just the vector $\hatr$
after an
infinitesimal rotation $R_{\hata}(\Delta\xi)$.
By applying successively
a large number of
infinitesimal rotations
$R_{\hata}(\Delta\xi)$, we get
a rotation
$R_{\hata}(\xi)$
over a finite angle $\xi\in\RR$:

\beq
e^{-i\frac{\xi}{2}\sigma_\hata}
\sigma_\hatr
e^{i\frac{\xi}{2}\sigma_\hata}
=
\sigma_{R_{\hata}(\xi)\hatr}\
\;.
\eeq

\section{A SEO That Takes $\ket{s'}$ to $\ket{t}$}
In this section,
we will explain a
formalism and accompanying
geometrical picture that
can be used to describe
both the original Grover's
algorithm, and the AFGA
proposed in this paper.

Our goal is to
find a SEO of
SU(2) transformations that
takes a starting state
$\ket{s'}\in\CC^2$
to a target state $\ket{t}\in \CC^2$.
Without loss of generality, we will take

\beq
\ket{t} = \ket{\hatz} = \ket{0}
\;
\eeq
and

\beq
\ket{s'}=\ket{\hats'}
\mbox{  where  }
\hats' = (\sin\gamma)\hatx + (\cos\gamma)\hatz
\;
\eeq
with
$\gamma\in [0,\pi]$.

Let $E_j$ for $j=0,1, \ldots,\jmax$
denote
\beq
E_j = e^{-i\frac{\xi_j}{2}\sigma_{\hata_j}}
\;
\eeq
where $\xi_j\in\RR$ and
the $\hata_j$ are unit $\RR^3$ vectors.
Suppose $\hats_{fin}$
is generated as follows

\beq
\hats_{fin} =
[\prod_{j=\jmax\rarrow 0}R_{\hata_j}(\xi_j)]
\hats'
\;.
\eeq
(The arrow in the subscript of the
product sign indicates the
order in which to
multiply the terms,
this being an ordered product.)
It follows that

\beqa
\left|\bra{0}
(\prod_{j=\jmax\rarrow 0}E_j)
\ket{s'}\right|^2
&=&
\bra{0}
(\prod_{j=\jmax\rarrow 0}E_j)
\ket{s'}\bra{s'}
(\prod_{j=0\rarrow \jmax}E^\dagger_j)
\ket{0}
\\&=&
\bra{0}
(\prod_{j=\jmax\rarrow 0}E_j)
\left(\frac{1+\sigma_{\hats'}}{2}\right)
(\prod_{j=0\rarrow \jmax} E^\dagger_j)
\ket{0}
\\&=&
\frac{1}{2}
\left(
1+
\bra{0}
(\prod_{j=\jmax\rarrow 0}
e^{-i\frac{\xi_j}{2}\sigma_{\hata_j}})
\sigma_{\hats'}
(\prod_{j=0\rarrow \jmax}
e^{i\frac{\xi_j}{2}\sigma_{\hata_j}})
\ket{0}
\right)
\\&=&
\frac{1}{2}
\left(
1+
\bra{0}
\sigma_{\hats_{fin}}
\ket{0}
\right)
\\&=&
\frac{1}{2}
\left(
1+
[\hats_{fin}]_z
\right)
\;.
\label{eq-pre-err}
\eeqa
If we define $ERR$
by

\beq
ERR =
1 - \left|\bra{0}
(\prod_{j=\jmax\rarrow 0}E_j)
\ket{s'}\right|^2
\;,
\label{eq-err-def}
\eeq
then Eq.(\ref{eq-pre-err}) implies that

\beq
ERR
=
\frac{1}{2}(1-[\hats_{fin}]_z)
\;.
\label{eq-err-sfin}
\eeq
$ERR\geq 0$ is a measure of error.
By Eq.(\ref{eq-err-def}), $ERR$
 decreases towards zero as the SEO takes
$\ket{s'}$ closer to $\ket{t}$.
Eq.(\ref{eq-err-sfin})
agrees with our expectation that $ERR$
goes to zero as the Z component
of $\hats_{fin}$ approaches one.

In the original Grover's algorithm,
$\ket{s'} = U\ket{s}$, where
$\ket{s}=\ket{0}^{\otimes\nb}$
and $U = H^{\otimes\nb}$.
(Here $H$ denotes the one-bit Hadamard matrix
$\frac{1}{\sqrt{2}}(\sigx + \sigz)$.
Also $\ket{t}=\ket{t_{\nb-1}}\otimes\ldots\otimes\ket{t_1}\otimes\ket{t_0}$
with $t_i=\delta_i^{i_0}$
for some $i_0\in \{0,1, \ldots, \nb-1\}$.
Furthermore,

\beq
\prod_{j=\jmax\rarrow 0}E_j = G^{N_{steps}}
\;,
\eeq
where $N_{steps}$ is the number of
steps (i.e., queries) and

\beqa
G &=& - (-1)^{\ket{s'}\bra{s'}}
(-1)^{\ket{t}\bra{t}}
\\&=&
e^{i\pi} e^{i \pi\nbar_{\hats'}} e^{i \pi\nbar_{\hatz}}
\\&=&
e^{i \frac{\pi}{2}\sigma_{\hats'}}
e^{i \frac{\pi}{2}\sigz}
\;.
\eeqa

Note that $G$ corresponds to a small rotation about
the $\haty$ axis. Indeed,

\beqa
G &=& -(-1)^{\nbar_{\hats'}}(-1)^{\nbar_\hatz}
\\&=&
-\sigma_{\hats'}\sigz
\\&=&
-(\hats'\cdot\hatz + i\sigma_{\hats'\times\hatz})
\\&=&
-(\cos\gamma -i\sigy\sin\gamma)
\\&=&
e^{i(\pi-\gamma)\sigy}=e^{i\frac{\Delta\gamma}{2}\sigy}
\;
\eeqa
where $\Delta\gamma=2(\pi-\gamma)\geq 0$.
Thus
\beq
R_{\hats'}(-\pi)R_{\hatz}(-\pi)
=R_{\haty}(-\Delta\gamma)
\;.
\eeq
From $\cos(\frac{\gamma}{2})=\av{s'|t}=\frac{1}{\sqrt{\ns}}$,
it follows that
$\Delta\gamma=\calo(\frac{1}{\sqrt{\ns}})$.

Note also that $G$ can be p-compiled trivially. Indeed,
if $P_0$ and $P_1$ are defined as in Eqs.(\ref{eq-p0-p1}),

\begin{subequations}\label{eq-compiling}
\beq
(-1)^{\ket{t}\bra{t}}
=(-1)^{\prod_{\beta=0}^{\nb-1}P_{t_i}(\beta)}
\;,
\eeq
and

\beq
(-1)^{\ket{s'}\bra{s'}}
=
H^{\otimes\nb}\left[
(-1)^{\ket{0^\nb}\bra{0^\nb}}
\right]H^{\otimes\nb}
=
H^{\otimes\nb}\left[
(-1)^{\prod_{\beta=0}^{\nb-1}P_{0}(\beta)}
\right]H^{\otimes\nb}
\;.
\eeq
\end{subequations}

Next, let's describe our
AFGA.
We take

\beqa
\prod_{j=\jmax\rarrow 0}E_j &=&
\prod_{j=\infty\rarrow 0}
(e^{i \alpha_j\ket{s'}\bra{s'}}
e^{i \Delta\lambda\ket{t}\bra{t}})
\\&=&
\prod_{j=\infty\rarrow 0}
(e^{\frac{i}{2}(\alpha_j+\Delta\lambda)}
e^{i \frac{\alpha_j}{2}\sigma_{\hats'}}
e^{i \frac{\Delta\lambda}{2}\sigz})
\;,
\eeqa
for some $\Delta\lam\in[0,\pi]$
and some infinite sequence of
$\alpha_j\in\RR$. In the original
Grover's algorithm, $\Delta\lam$
and all the $\alpha_j$ are fixed
at $\pi$. That's why we say
our algorithm is adaptive.
We will also assume that

\beq
R_{\hats'}(-\alpha_j)R_{\hatz}(-\Delta\lambda)
=R_\haty(-\Deltabar\gamma_j)
\;
\eeq
for some $\Deltabar\gamma_j\in \RR$, and
that
 we know how to p-compile
$e^{i \alpha_j\ket{s'}\bra{s'}}$
and
$e^{i \Delta\lambda\ket{t}\bra{t}}$.
If $\ket{s'}$ and $\ket{t}$
are the same as in the original Grover's
algorithm, then we do indeed know
how to p-compile these operators. (Just replace
the $(-1)=e^{i\pi}$ phase factors
in Eqs.(\ref{eq-compiling}) by $e^{i\alpha_j}$
and $e^{i\Delta\lambda}$.)

Our AFGA has a total
of two input parameters: $\Delta\lam\in[0,\pi]$
(the angle
of each consecutive $\hatz$ rotation), and
$\gamma\in[0,\pi]$ (the angle which $\hats'$
makes with $\hatz$). For the AFGA
to be fully specified, we still need
to specify, as a function
of these two input parameters, a suitable sequence
of $\alpha_j$ that makes $ERR$ converge to zero.
We will do this in the next section.

\section{Bouncing Between Two Longitudes}
In this section, we
give a suitable
sequence $\{\alpha_j\}_{j=0}^\infty$
for our AFGA, as a function of the
two input parameters $\Delta\lam$
and $\gamma$.
We will do this guided
by the geometrical picture
of bouncing between
two great circles of longitude
of the unit sphere.

\begin{figure}[h]
    \begin{center}
    \epsfig{file=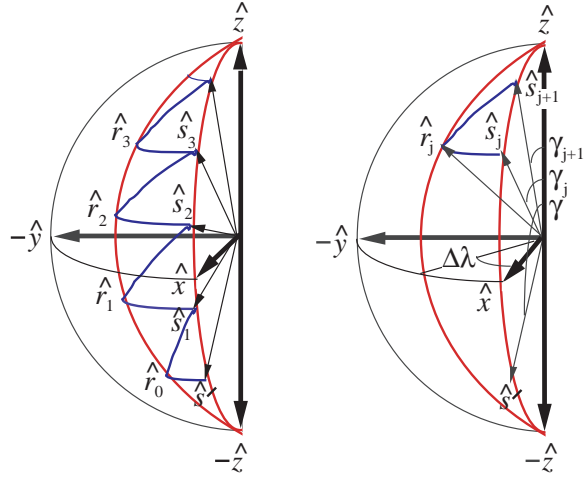, height=2.7in}
    \caption{Unit $\RR^3$ vectors and angles
    used in our AFGA. Note how we ``bounce between
    two longitudes".}
    \label{fig-bouncing}
    \end{center}
\end{figure}

\begin{figure}[h]
    \begin{center}
    \epsfig{file=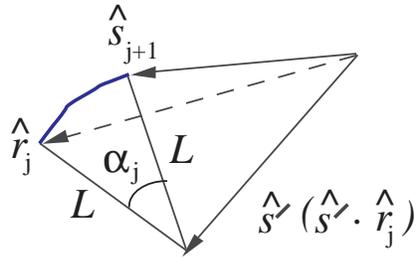, height=1.4in}
    \caption{Geometry defining angle $\alpha_j$
    and radius $L$. }
    \label{fig-alpha-j}
    \end{center}
\end{figure}

It is convenient
to define two sequences
of unit $\RR^3$ vectors $\{\hats_j\}_{j=0}^\infty$,
and $\{\hatr_j\}_{j=0}^\infty$
by the recursion relation:

\beq
\hats_0 = \hats'
\;
\eeq

\beq
\begin{array}{l}
\hatr_j = R_\hatz(-\Delta\lam)\hats_j,
\\
\hats_{j+1} = R_{\hats'}(-\alpha_j)\hatr_j
\\
\hats_{j+1}= R_\haty(-\Deltabar\gamma_j)\hats_j
\end{array}
\;
\label{eq-s-r-rots}
\eeq
for $j=0,1,\ldots$.
According to Eqs.(\ref{eq-s-r-rots}),
\beq
\begin{array}{l}
\angle(\hatr_j, \hats_j)=\Delta\lam,
\\
\angle(\hats_{j+1},\hatr_j) = \alpha_j,
\\
\angle(\hats_{j+1},\hats_j)= \Deltabar\gamma_j
\end{array}
\;.
\eeq

See Figs.\ref{fig-bouncing}
for a geometrical picture of
the $\hats_j$ and $\hatr_j$ sequences of vectors.
The arrowheads of all
the $\hats_j$
vectors lie on the great circle of longitude
located at the intersection of
the $\hatx-\hatz$ plane and the unit sphere.
The arrowheads of all
the $\hatr_j$
vectors lie on the great circle of longitude
which makes an angle of $\Delta\lam$
with the $\hats_j$ great circle of longitude.

The angles $\gamma_j$ and $\Deltabar\gamma_j$
are defined in terms of the vectors
$\hats_j$ as follows:

\beq
\angle(\hats',\hatz)=\gamma = \gamma_0
\;
\eeq
and

\beq
\angle(\hats_j,\hatz)=\gamma_j
\;,\;\;
\angle(\hats_{j+1},\hatz)=\gamma_{j+1}
\;\;
\Deltabar\gamma_j = -\Delta\gamma_j = \gamma_j - \gamma_{j+1}
\;
\eeq
for $j=0,1,\ldots$.

Note from Fig.\ref{fig-bouncing}
and \ref{fig-alpha-j} that

\beq
\angle(\hatr_j,\hats') =
\angle(\hats_{j+1},\hats')=
\gamma-\gamma_{j+1}=
\gamma-\gamma_{j}+\Deltabar\gamma_j
\;.
\label{eq-angle-rjs}
\eeq

Since
\beqa
\hatr_j &=& R_\hatz(-\Delta\lam)\hats_j
\\&=&
\hatz(\hatz\cdot\hats_j)
-\hatz\times\hats_j S_{\Delta\lam}
+[\hats_j - (\hats_j\cdot\hatz)\hatz]C_{\Delta\lam}
\\&=&
\hatz C_{\gamma_j}(1- C_{\Delta\lam})
-\haty S_{\gamma_j}S_{\Delta\lam}
+\hats_j C_{\Delta\lam}
\;,
\label{eq-rj}
\eeqa
it follows that

\beqa
\hatr_j\cdot\hats'&=&
C_{\gamma}C_{\gamma_j}(1-C_{\Delta\lam})
+ \cos(\gamma-\gamma_j)C_{\Delta\lam}
\\&=&
C_{\gamma}C_{\gamma_j}
+ S_\gamma S_{\gamma_j}C_{\Delta\lam}
\;.
\eeqa

As is usual in the C programming language,
define $\theta = \atan 2(y,x)$
only if $\tan\theta = y/x$.(Think of the comma
in $\atan 2(y,x)$ as a slash indicating division).
Eq.(\ref{eq-angle-rjs}) implies that

\beq
\gamma-\gamma_j+\Deltabar\gamma_j=
{\atan 2}(\pm \sqrt{1-(\hatr_j\cdot\hats')^2},
\hatr_j\cdot\hats')
\;.
\label{eq-pm}
\eeq
Assume
$\angle(\hatr_j,\hats')\in[0,\pi]$.
This means we choose the
solution with the positive sign
in Eq.(\ref{eq-pm}).

To summarize,
we've shown that

\begin{subequations}\label{eq-box-dgj}
\fbox{\parbox{5.5in}{
\beq
\Deltabar\gamma_j=-\gamma+\gamma_j+
{\atan 2}(\sqrt{1-(\hatr_j\cdot\hats')^2},
\hatr_j\cdot\hats')
\;
\eeq
}}
where

\fbox{\parbox{5.5in}{
\beq
\hatr_j\cdot\hats'=
C_{\gamma}C_{\gamma_j}
+ S_\gamma S_{\gamma_j}C_{\Delta\lam}
\;
\eeq
}}
\end{subequations}
Eqs.(\ref{eq-box-dgj}) and $\gamma_0=\gamma$ allows us to find
the sequence $\{\gamma_j\}_{j=0}^\infty$.
The
sequence $\{\gamma_j\}_{j=0}^\infty$
 and $\hats_0=\hats'$ allows us to find
the sequence $\{\hats_j\}_{j=0}^\infty$.

Next we solve for the
sequence $\{\alpha_j\}_{j=0}^\infty$,
assuming that we know the two
input parameters $\Delta\lam$ and
$\gamma$, and the
sequence $\{\gamma_j\}_{j=0}^\infty$.
We can find $\alpha_j$
by considering
Fig.\ref{fig-alpha-j}.
That figure defines $L$ as

\beq
L = |\hatr_j - \hats'(\hats'\cdot\hatr_j)|
=\sqrt{1-(\hatr_j\cdot\hats')^2}
=|\sin(\gamma-\gamma_j+\Deltabar\gamma_j)|
=\sin(\gamma-\gamma_j+\Deltabar\gamma_j)
\;.
\eeq
Note also that

\beq
\hats_{j+1}=
R_{\haty}(-\Deltabar\gamma_j)\hats_j=
-\haty\times\hats_j S_{\Deltabar\gamma_j}
+ \hats_j C_{\Deltabar\gamma_j}
\;.
\eeq

It follows from Fig.\ref{fig-alpha-j}
that

\beqa
L^2 \sin\alpha_j &=&
[\hatr_j - \hats'(\hats'\cdot\hatr_j)]\times
[\hats_{j+1} - \hats'(\hats'\cdot\hatr_j)]\cdot(-\hats')
\\&=&
\hatr_j\times\hats_{j+1}\cdot(-\hats')
\\&=&
S_{\Deltabar\gamma_j}[\hatr_j\times(\haty\times\hats_j)\cdot\hats']
+C_{\Deltabar\gamma_j}[-\hatr_j\times\hats_j\cdot\hats']
\;.
\eeqa
But by
virtue of Eq.(\ref{eq-rj}) for $\hatr_j$,

\beq
\hatr_j\times(\haty\times\hats_j)\cdot\hats'
=
-(\hats_j\cdot\hats')(\hatr_j\cdot\haty)
=
\cos(\gamma-\gamma_j)S_{\gamma_j}S_{\Delta \lam}
\;,
\eeq
and

\beq
\hatr_j\times\hats_j\cdot\hats'
=
-(\haty\times\hats_j\cdot\hats')S_{\gamma_j}S_{\Delta \lam}
=
-\sin(\gamma-\gamma_j)S_{\gamma_j}S_{\Delta \lam}
\;
\eeq
so

\fbox{\parbox{5.5in}{
\beq
L^2\sin\alpha_j=
\sin(\gamma -\gamma_j + \Deltabar\gamma_j)
S_{\gamma_j}S_{\Delta \lam}
\;.
\label{eq-sine-alpj}
\eeq
}}

\vspace{1em}
It also follows from
Fig.\ref{fig-alpha-j} that

\beqa
L^2\cos\alpha_j &=&
[\hatr_j - \hats'(\hats'\cdot\hatr_j)]\cdot
[\hats_{j+1} - \hats'(\hats'\cdot\hatr_j)]
\\&=&
[\hatr_j - \hats'(\hats'\cdot\hatr_j)]\cdot
\hats_{j+1}
\\&=&
[\hatr_j - \hats'(\hats'\cdot\hatr_j)]\cdot
[-\haty\times\hats_j S_{\Deltabar\gamma_j}
+ \hats_j C_{\Deltabar\gamma_j}]
\\&=&
\left\{\begin{array}{l}
S_{\Deltabar\gamma_j}
[\hatr_j\times\hats_j\cdot\haty +
\sin(\gamma-\gamma_j)\hatr_j\cdot\hats']
\\
+C_{\Deltabar\gamma_j}
[\hatr_j\cdot\hats_j
-\cos(\gamma-\gamma_j)\hatr_j\cdot\hats']
\end{array}
\right.
\;.
\eeqa
But by
virtue of Eq.(\ref{eq-rj}) for $\hatr_j$,

\beq
\hatr_j\times\hats_j\cdot\haty=
(\hatz\times\hats_j\cdot\haty) C_{\gamma_j}(1-C_{\Delta\lam})
=
S_{\gamma_j}C_{\gamma_j}(1-C_{\Delta\lam})
\;,
\eeq
and

\beq
\hatr_j\cdot\hats_j =
C^2_{\gamma_j}(1-C_{\Delta\lam})
 + C_{\Delta\lam}=
C^2_{\gamma_j} + S^2_{\gamma_j}C_{\Delta\lam}
\;
\eeq
so

\fbox{\parbox{5.5in}{
\beq
L^2\cos\alpha_j =
\left\{\begin{array}{l}
S_{\Deltabar\gamma_j}
[C_{\gamma_j}S_{\gamma_j}(1-C_{\Delta\lam}) +
\sin(\gamma-\gamma_j)\hatr_j\cdot\hats']
\\
+C_{\Deltabar\gamma_j}
[C^2_{\gamma_j} + S^2_{\gamma_j}C_{\Delta\lam}
-\cos(\gamma-\gamma_j)\hatr_j\cdot\hats']
\end{array}
\right.
\;.
\label{eq-cos-alpj}
\eeq
}}

\vspace{1em}
To summarize,

\fbox{\parbox{5.5in}{
\beq
\alpha_j=\atan2(L^2\sin\alpha_j, L^2 \cos\alpha_j)
\;,
\eeq
}}
where
$L^2\sin\alpha_j$ and $L^2 \cos\alpha_j$
are given by Eqs.(\ref{eq-sine-alpj})
and (\ref{eq-cos-alpj}).

\section{Software and Numerical Results}
In this section, we describe some
simple software
that calculates,
among other things, the phases $\alpha_j$ used
in our AFGA. We also
present and discuss some examples
of the output of the software.
\begin{figure}[h]
    \begin{center}
{\tiny
\begin{verbatim}
gamma(degs) = 1.7315e+02
del_lam(degs) = 1.3500e+02
num_steps = 20
j   gam_j(degs) alp_j(degs) vr_x        vr_y        vr_z        vs_x        vs_y        vs_z
0   1.7315e+02  1.5735e+02  -8.4337e-02 -8.4337e-02 -9.9286e-01 1.1927e-01  0.0000e+00  -9.9286e-01
1   1.6050e+02  1.4576e+02  -2.3607e-01 -2.3607e-01 -9.4263e-01 3.3385e-01  -2.7756e-17 -9.4263e-01
2   1.4835e+02  1.4171e+02  -3.7109e-01 -3.7109e-01 -8.5122e-01 5.2480e-01  -1.1102e-16 -8.5122e-01
3   1.3636e+02  1.3947e+02  -4.8795e-01 -4.8795e-01 -7.2375e-01 6.9006e-01  -1.6653e-16 -7.2375e-01
4   1.2448e+02  1.3795e+02  -5.8289e-01 -5.8289e-01 -5.6611e-01 8.2433e-01  -2.2204e-16 -5.6611e-01
5   1.1266e+02  1.3676e+02  -6.5253e-01 -6.5253e-01 -3.8523e-01 9.2282e-01  -1.1102e-16 -3.8523e-01
6   1.0089e+02  1.3572e+02  -6.9438e-01 -6.9438e-01 -1.8888e-01 9.8200e-01  5.5511e-17  -1.8888e-01
7   8.9160e+01  1.3472e+02  -7.0703e-01 -7.0703e-01 1.4652e-02  9.9989e-01  1.6653e-16  1.4652e-02
8   7.7476e+01  1.3369e+02  -6.9028e-01 -6.9028e-01 2.1686e-01  9.7620e-01  -5.5511e-17 2.1686e-01
9   6.5834e+01  1.3253e+02  -6.4514e-01 -6.4514e-01 4.0938e-01  9.1236e-01  -2.2204e-16 4.0938e-01
10  5.4242e+01  1.3107e+02  -5.7381e-01 -5.7381e-01 5.8436e-01  8.1149e-01  0.0000e+00  5.8436e-01
11  4.2712e+01  1.2901e+02  -4.7964e-01 -4.7964e-01 7.3478e-01  6.7831e-01  -2.2204e-16 7.3478e-01
12  3.1268e+01  1.2557e+02  -3.6702e-01 -3.6702e-01 8.5475e-01  5.1905e-01  -1.6653e-16 8.5475e-01
13  1.9971e+01  1.1787e+02  -2.4151e-01 -2.4151e-01 9.3986e-01  3.4155e-01  -3.3307e-16 9.3986e-01
14  9.0040e+00  8.5904e+01  -1.1067e-01 -1.1067e-01 9.8768e-01  1.5650e-01  1.9429e-16  9.8768e-01
15  -4.8000e-01 -2.7100e+00 5.9237e-03  5.9237e-03  9.9996e-01  -8.3774e-03 7.5114e-16  9.9996e-01
16  3.4738e-01  2.1347e+00  -4.2871e-03 -4.2871e-03 9.9998e-01  6.0629e-03  -7.6762e-16 9.9998e-01
17  -2.4109e-01 -1.3945e+00 2.9753e-03  2.9753e-03  9.9999e-01  -4.2078e-03 4.4149e-16  9.9999e-01
18  1.7254e-01  1.0412e+00  -2.1293e-03 -2.1293e-03 1.0000e+00  3.0113e-03  5.0394e-16  1.0000e+00
19  -1.2090e-01 -7.0794e-01 1.4921e-03  1.4921e-03  1.0000e+00  -2.1101e-03 -1.6948e-15 1.0000e+00
20  8.6014e-02  5.1447e-01  -1.0615e-03 -1.0615e-03 1.0000e+00  1.5012e-03  1.1501e-15  1.0000e+00
\end{verbatim}
}
\caption{Typical output
    produced by running afga.m.}
    \label{fig-afga-out}
    \end{center}
\end{figure}

We've written 3 Octave/Matlab m-files
called {\tt afga.m}, {\tt afga\_step.m}
and {\tt afga\_rot.m}
that implement some of the results of this paper. The
main file {\tt afga.m}
calls functions in
{\tt afga\_step.m} and {\tt afga\_rot.m}.

The first 3 lines of {\tt afga.m}
instantiate the 3 input parameters
{\tt g0\_degs} (=$\gamma$ in degrees),
{\tt del\_lam\_degs}($=\Delta\lam$
in degrees),
and {\tt num\_steps}
(= maximum value of $j$
that will be considered. $j$ will
range from 0 to {\tt num\_steps})

Each time
{\tt afga.m} runs
successfully, it outputs
a text file called
{\tt afga.txt}. Fig.\ref{fig-afga-out}
illustrates a typical {\tt afga.txt} file.
The first 3 lines record
the inputs.
The next line
labels the columns of the file.
The column labels are
\begin{itemize}
\item {\tt j} = $j$
\item {\tt gam\_j(degs)} = $\gamma_j$ in degrees
\item {\tt alp\_j(degs)} = $\alpha_j$ in degrees
\item {\tt vr\_x} = $[\hatr_j]_x$
\item {\tt vr\_y} = $[\hatr_j]_y$
\item {\tt vr\_z} = $[\hatr_j]_z$
\item {\tt vs\_x} = $[\hats_j]_x$
\item {\tt vs\_y} = $[\hats_j]_y$
\item {\tt vs\_z} = $[\hats_j]_z$
\end{itemize}
Following the line of column labels
are {\tt num\_steps}+1 lines of
output data.

In {\tt afga.txt},
data in each line
is separated by a tab. Thus, the full
{\tt afga.txt} file can be
cut-and-pasted into an Excel spreadsheet
or other plotting software
in order to plot it.

\begin{figure}[t]
    \begin{center}
    \epsfig{file=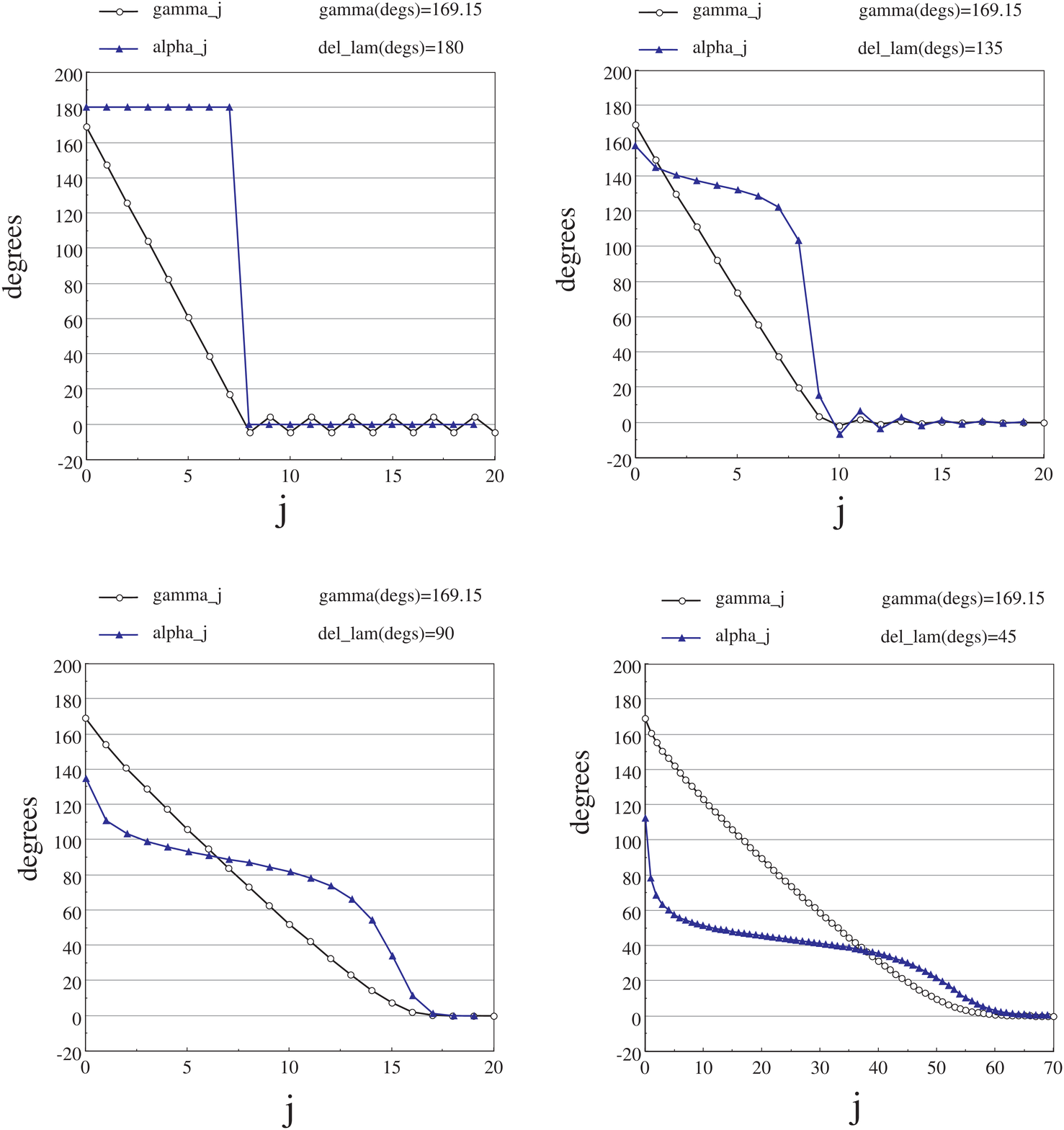, height=5in}
    \caption{Values of $\gamma_j$
    and $\alpha_j$ obtained with {\tt afga.m}
    for $\gamma=169.15^0$ and various
    values of $\Delta\lam$}
    \label{fig-g170}
    \end{center}
\end{figure}
\mbox{ }
\newpage\mbox{ }
\begin{figure}[h]
    \begin{center}
    \epsfig{file=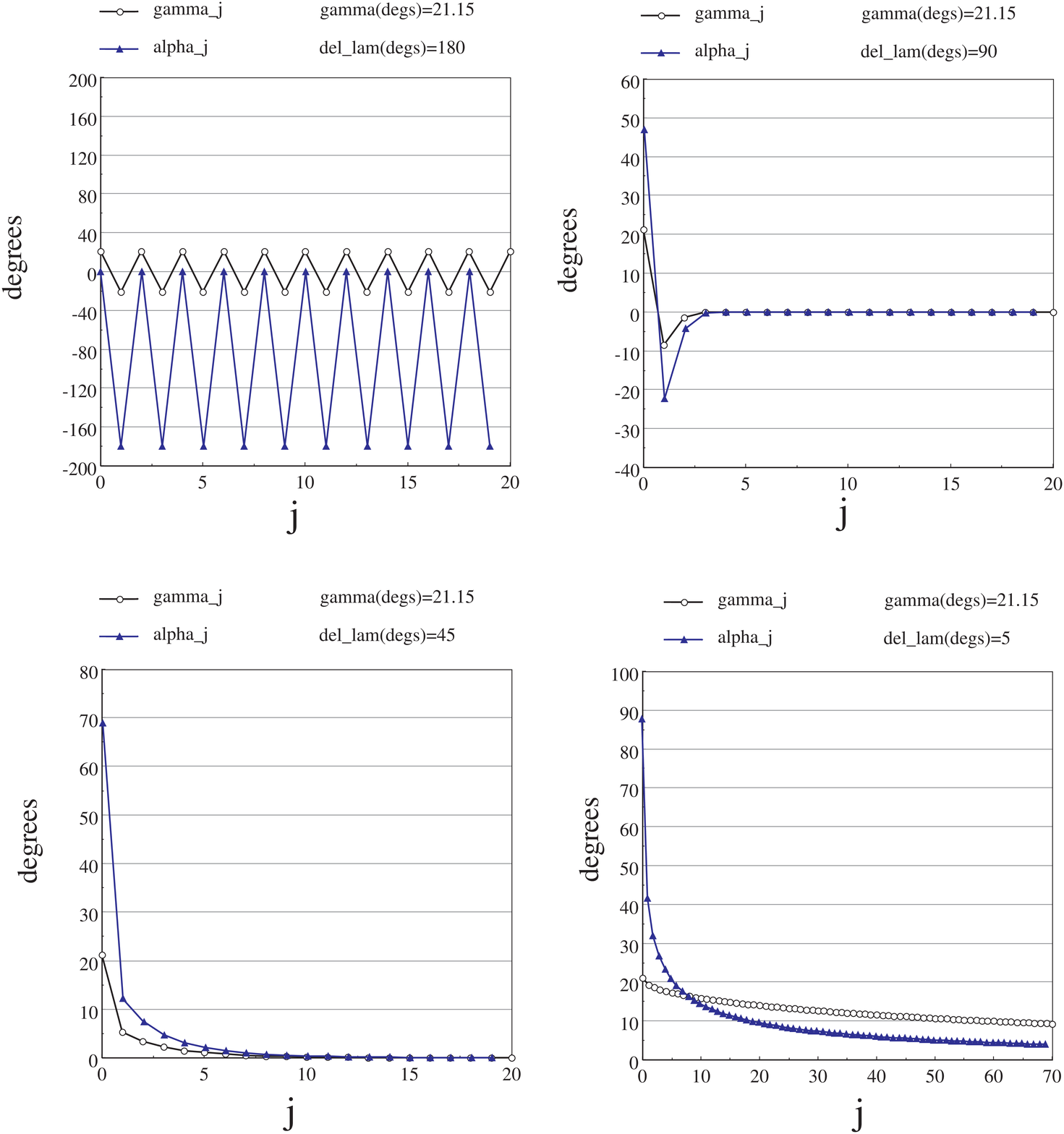, height=5in}
    \caption{Values of $\gamma_j$
    and $\alpha_j$ obtained with {\tt afga.m}
    for $\gamma=21.15^0$ and various
    values of $\Delta\lam$}
    \label{fig-g20}
    \end{center}
\end{figure}

Fig.\ref{fig-g170}
shows the values of $\gamma_j$
and $\alpha_j$ obtained with {\tt afga.m}
for $\gamma=169.15^0$ and various
values of $\Delta\lam$.
Fig.\ref{fig-g20}
shows the same thing
but
for $\gamma=21.15^0$.
We see that $\gamma_j$
decreases
almost linearly
from $\gamma$
to near zero.
The behavior of $\gamma_j$
near
zero
depends on the value
of $\Delta\lam$.
For $0\leq \Delta\lam<\pi$,
$\gamma_j$ goes to zero
without too many oscillations.
For $\Delta\lam$ precisely
equal to $\pi$,
$\gamma_j$ never
converges to zero.
It gets trapped near zero,
oscillating about it
with a constant amplitude.

In Appendix \ref{app-del_lam_is_pi},
we discuss in more detail
 the behavior of our AFGA
when $\Delta\lam=\pi$.
This case most closely resembles
the original Grover's algorithm.

In Appendix \ref{app-continuum},
we discuss the continuum
limit where $\Deltabar\gamma_j$
tends to zero for all $j$. This
limit is a smoothed out version
of the discrete case.
It is easily solved, and gives
a good idea of
the rate of convergence
of $\gamma_j$
(and of $ERR$) towards zero
(when $\Delta\lam\neq \pi$) as
$j$ tends to infinity.

\appendix

\section{Appendix: When $\Delta\lam=\pi$}
\label{app-del_lam_is_pi}\begin{figure}[h]
    \begin{center}
    \epsfig{file=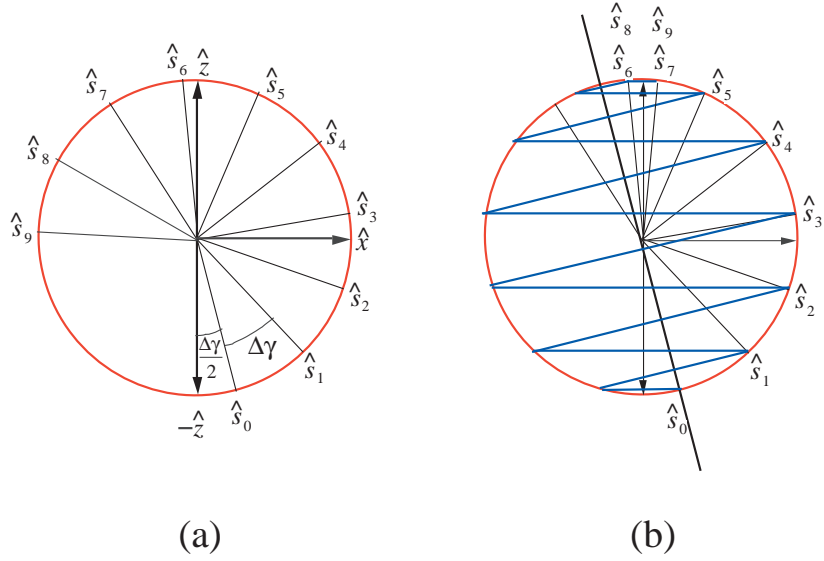, height=3in}
    \caption{(a)$\hats_j$ vectors for
    original Grover's algorithm.
    (b)$\hats_j$ vectors for
    our AFGA
    with $\Delta\lam=\pi$
    and the same $\gamma$ as in (a).}
    \label{fig-grover-ada}
    \end{center}
\end{figure}

\begin{figure}[h]
    \begin{center}
    \epsfig{file=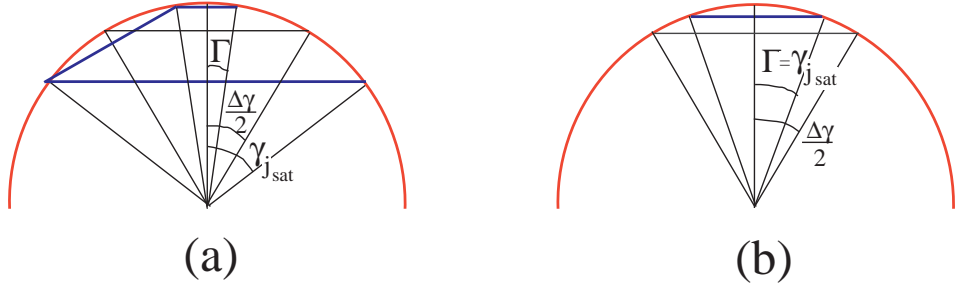, height=1.5in}
    \caption{(a)$\Gamma$ when
    $\gamma_{\jsat}>\frac{\Delta\gamma}{2}$.
    (b)$\Gamma$ when
    $\gamma_{\jsat}<\frac{\Delta\gamma}{2}$.}
    \label{fig-ada}
    \end{center}
\end{figure}
In this section, we
discuss the $\Delta\lam=\pi$ case of our AFGA.

Fig.\ref{fig-grover-ada}(a)
shows the pattern of the $\hats_j$ vectors
for the original Grover's algorithm,
and Fig.\ref{fig-grover-ada}(b)
shows the pattern for our AFGA with $\Delta\lam=\pi$
and the same $\gamma$ as in (a).
There is a critical $j$, call
it $\jsat$ (``sat" for saturation).
(a) and (b) have the same $\hats_j$ vectors
for $j=0,1,\ldots, \jsat$.
For $j>\jsat$,
the $\hats_j$ vectors of (a)
continue to decrease their angle (with respect to $\hatz$)
at a uniform rate, past the North Pole, whereas
the $\hats_j$ vectors of (b) get trapped
in the neighborhood
of the North Pole, bouncing back and forth,
making an angle of $\pm\Gamma$
with respect to $\hatz$. In other words,
the pattern observed is like this:

\beq
\begin{array}{c|c|c}
j&\gamma_j\mbox{ for }(a)&\gamma_j\mbox{ for }(b)\\
\hline\hline
\vdots & \vdots & \vdots \\\hline
\jsat-1 & \gamma_{\jsat}+ \Delta\gamma& \gamma_{\jsat}+\Delta\gamma \\\hline
\jsat   & \gamma_{\jsat}              & \gamma_{\jsat}              \\\hline
\jsat+1 & \gamma_{\jsat}- \Delta\gamma& -\Gamma                       \\\hline
\jsat+2 & \gamma_{\jsat}-2\Delta\gamma&  \Gamma                       \\\hline
\jsat+3 & \gamma_{\jsat}-3\Delta\gamma& -\Gamma                       \\\hline
\jsat+3 & \gamma_{\jsat}-4\Delta\gamma&  \Gamma                       \\\hline
\vdots & \vdots & \vdots\\\hline
\end{array}
;.
\eeq
$\jsat$ is defined by the constraints
that
$\gamma_{\jsat-1}>\Delta\gamma$ and
$0\leq\gamma_{\jsat}<\Delta\gamma$.

As shown by Fig.\ref{fig-ada},

\beqa
\Gamma &=&
\left\{
\begin{array}{ll}
\gamma_\jsat &\mbox{ if }
0\leq\gamma_\jsat\leq\frac{\Delta\gamma}{2}\\
\Delta\gamma-\gamma_\jsat &\mbox{ if }
\frac{\Delta\gamma}{2}\leq\gamma_\jsat\leq\Delta\gamma
\end{array}
\right.
\\
&=&\min(\gamma_\jsat,
\Delta\gamma-\gamma_\jsat)
\;.
\label{eq-Gamma-formula}
\eeqa
For example,

\beq
\begin{array}{r||r|r|r}
\gamma\mbox{(degs)} & \Delta\gamma\mbox{(degs)} & \gamma_\jsat\mbox{(degs)} & \Gamma\mbox{(degs)}\\\hline\hline
160 & 2(20)& 0 & 0\\\hline
164 & 2(16)& 4 & 4\\\hline
166 & 2(14)& 26 & 2\\\hline
\end{array}
\;.
\label{eq-Gamma-example}
\eeq
The last column of Eq.(\ref{eq-Gamma-example})
was calculated using Eq.(\ref{eq-Gamma-formula}) and
then checked using {\tt afga.m}.

\section{Appendix: Continuum Limit}
\label{app-continuum}

\begin{figure}[h]
    \begin{center}
    \epsfig{file=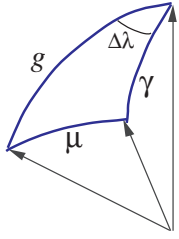, height=1.2in}
    \caption{A spherical triangle with sides
    of length $\mu$, $\gamma$ and $g$,
    and angle $\Delta\lam$ between the $g$ and $\gamma$
    sides.
    The spherical triangle has sides which are
    segments of great circles of the unit sphere.}
    \label{fig-sph-triangle}
    \end{center}
\end{figure}

In this section, we explore $\gamma_j$ in the
continuum limit.

Suppose we take the limit
where $\Deltabar\gamma_j$
tends to zero for all $j\geq 0$.
We replace
$j$ by a real number $t\geq 0$
and $\gamma_j$ by a continuous
function $g(t)$ of $t$.
$\Deltabar\gamma_j = \gamma_j -\gamma_{j+1}\rarrow
-\frac{dg}{dt}$ and Eq.(\ref{eq-box-dgj}) tends to

\beq
-\frac{dg}{dt} = -\gamma + g +\atan2(|S_{\mu(g)}|,C_{\mu(g)})
\;,
\label{eq-dg-dt}
\eeq
where $\mu(g)$ satisfies

\beq
C_{\mu(g)} = C_\gamma C_g + S_\gamma S_g C_{\Delta\lam}
\;.
\label{eq-cmu-def}
\eeq

%Note that, contrary to the discrete case,
%in the continuum limit
%there are no oscillations
%of $g$ about its zero value.
%If there were, there would
%be a countable set
%of points $g$ at which
%$\frac{dg}{dt}=0$ was discontinuous.

Henceforth, we will restrict our attention to the case
$0\leq g\leq \gamma\leq \pi$.

In general, note that if
$\cos(\alpha)=\cos(\beta)$, then
$\alpha = \pm\beta + 2\pi N$ for some integer $N$.
Hence, Eq.(\ref{eq-cmu-def}) doesn't
specify $\mu(g)$ uniquely.
However, the ``Law of Cosines" of
spherical trigonometry tells us that
one possible
value for $\mu(g)$
is the length of the side of the spherical
triangle portrayed in Fig.\ref{fig-sph-triangle}.
Henceforth, we will
identify $\mu(g)$
with this unique geometrical value.
When $\mu$ is given this geometrical value, since $0\leq g \leq \gamma\leq \pi$,
$\mu\in[0,2\pi]$.
Since $\atan2(|S_\mu|,C_\mu)\in[0,\pi]$,
we must have

\beqa
\atan2(|S_\mu|,C_\mu)&=&
\left\{
\begin{array}{ll}
\mu&\mbox{ if }\mu\in[0,\pi]
\\
2\pi-\mu&\mbox{ if }\mu\in[\pi,2\pi]
\end{array}\right.
\\&=&
\min(\mu, 2\pi-\mu)
\;.
\eeqa
Hence, when $\mu(g)$
is given its geometrical value,

\beq
-\frac{dg}{dt} = -\gamma + g +\min(\mu, 2\pi-\mu)
\;.
\label{eq-geom-dg-dt}
\eeq

Eq.(\ref{eq-geom-dg-dt})
can be solved in closed form
in some special cases:

\begin{enumerate}
\item[(a)]$|g|<<1$: In this case,
we get from Eq.(\ref{eq-cmu-def}),

\beq
C_\mu \approx C_\gamma + S_\gamma g C_{\Delta\lam}
\approx \cos(\gamma - gC_{\Delta\lam})
\;.
\eeq
so $\mu \approx \gamma - gC_{\Delta\lam}$. Hence,

\beq
-\frac{dg}{dt} \approx g
(1 - C_{\Delta\lam})
 \;\;\;\;\Rightarrow g\approx(const.)e^{-t(1- C_{\Delta\lam})}
\;.
\eeq

\item[(b)]
$0\leq\gamma-g<<1$: In this case,

\beq
-\frac{dg}{dt} \approx \min[\mu(\gamma), 2\pi-\mu(\gamma)]
\;\;\;\;
\Rightarrow g\approx\gamma - t\min[\mu(\gamma), 2\pi-\mu(\gamma)]
\;
\eeq

\begin{enumerate}
\item[(b.1)]$\Delta\lam = \pi$: In this case,
we see from Fig.\ref{fig-sph-triangle}
that $\mu(\gamma) = 2\gamma$. Hence

\beq
g\approx \left\{
\begin{array}{ll}
\gamma - t2(\pi-\gamma)=\gamma - t\Delta\gamma
& \mbox{if }\;\gamma>\frac{\pi}{2}\\
\gamma - t 2 \gamma
& \mbox{if }\;\gamma<\frac{\pi}{2}
\end{array}
\right.
\;.
\eeq
\item[(b.2)]$0\leq \Delta\lam << 1$:
In this case, by virtue of Eq.(\ref{eq-cmu-def}),

\beq
C_\mu\approx C_\gamma^2 + S_\gamma^2 C_{\Delta\lam}
\approx 1 - \frac{1}{2} S_\gamma^2 (\Delta\lam)^2
\approx \cos(S_\gamma\Delta\lam)
\;,
\eeq
so $\mu(\gamma) \approx S_\gamma \Delta\lam$. Hence

\beq
g\approx\gamma - tS_\gamma\Delta\lam
\;.
\eeq
\end{enumerate}
\end{enumerate}

And what is the maximum $|dg/dt|$
for $g\approx \gamma$?
When $g\approx \gamma$,

\beq
\max_{\Delta\lam\in[0,\pi]}|dg/dt|=
|dg/dt|_{\Delta\lam =\pi}=
\min[2\gamma, 2\pi - 2\gamma]
\;.
\eeq

\end{document}